# Characterizing Continuous and Discrete Hybrid Latent Spaces for Structural Connectomes

Gaurav Rudravaram[a], Lianrui Zuo[a], Adam M. Saunders[a], Michael E. Kim[b], Praitayini Kanakaraj[b], Nancy R. Newlin[b], Aravind R. Krishnan[a], Elyssa M. McMaster[a], Chloe Cho[c], Susan M. Resnick[d], Lori L. Beason Held[d], Derek Archer[e], Timothy J. Hohman[e], Daniel C. Moyer[b], and Bennett A. Landman[a,b,c]

[a]Department of Electrical and Computer Engineering, Vanderbilt University, Nashville, TN, USA; [b]Department of Computer Science, Vanderbilt University, Nashville, TN, USA; [c]Deparment of Biomedical Engineering, Vanderbilt University, Nashville, TN, USA; [d]Laboratory of Behavioral Neuroscience, National Institute on Aging, National Institutes of Health, Baltimore, MD, USA; [e]Vanderbilt University Medical Center, Vanderbilt Memory and Alzheimer's Center, Nashville, TN, USA

## ABSTRACT

Structural connectomes are detailed graphs that map how different brain regions are physically connected, offering critical insight into aging, cognition, and neurodegenerative diseases. However, these connectomes are high-dimensional and densely interconnected, which makes them difficult to interpret and analyze at scale. While low-dimensional spaces like PCA and autoencoders are often used to capture major sources of variation, their latent spaces are generally continuous and cannot fully reflect the mixed nature of variability in connectomes---which often include both continuous (e.g., connectivity strength) and discrete factors (e.g., imaging site). Motivated by this, we propose a variational autoencoder (VAE) with a hybrid latent space that jointly models the discrete and continuous components. We analyze a large dataset of 5,761 connectomes from 6 Alzheimer's disease studies with 10 unique acquisition protocols. Each connectome represents a single scan from a unique subject (3579 females, 2182 males), aged 22 to 102, with 4338 cognitively normal, 809 with mild cognitive impairment (MCI), and 614 with Alzheimer's disease (AD). Each connectome contains 121 brain regions defined by the BrainCOLOR atlas. We train our hybrid VAE in an unsupervised way and study what each component captures. We find that the discrete space is particularly effective at capturing subtle site-related differences, achieving an Adjusted Rand Index (ARI) of 0.65 with site labels, significantly outperforming traditional methods like PCA and standard VAE followed by clustering ($p \ll 0.05$). These results demonstrate that the hybrid latent space can disentangle distinct sources of variability in connectomes in an unsupervised manner, offering potential for large-scale connectome analysis.

## 1. INTRODUCTION

White matter abnormalities have been implicated in multiple neurodegenerative diseases such as Alzheimer's[1], mild traumatic brain injuries[2], bipolar disorder[3] as well as cognitive decline and aging[4]. Early changes in white matter have been associated with the onset of Alzheimer's disease (AD) and contribute significantly to cognitive deterioration[5]. As such, we must understand how the brain's structural organization changes with age, cognitive impairment, and disease progression to fully grasp the physical mechanisms of cognitive decline. While macrostructural alterations related to brain volume have been well documented, growing evidence also points to more subtle microstructural changes that occur both with aging and in the early stages of disease[6]. Investigating the dynamics of these microstructural change, and how they differ across conditions and across the lifespan, is essential for deepening our understanding of brain aging and neurodegeneration.

Diffusion-weighted magnetic resonance imaging (dMRI) offers a powerful, non-invasive approach to study white matter changes in the brain across the lifespan and in diverse clinical populations[7]. dMRI provides insight into the microstructural organization of white matter[8] by measuring the diffusion of water molecules in tissue. This modality yields scalar metrics such as fractional anisotropy (FA) and mean diffusivity (MD), which are commonly used to assess white matter integrity[9,10]. Beyond these scalar measures, dMRI can also be used to estimate the structural connectivity between brain regions through a process called tractography, which reconstructs the pathways of white matter fibers[11]. Following whole-brain tractography, we can summarize the complex brain network as a matrix defined by a number of regions-of-interest and the number of streamlines, or virtual projections of white matter pathways, that connect each pair of regions[12].

Structural connectomes derived from dMRI have been widely used to study brain organization in healthy controls and disease populations[13]. Alterations in connectome architecture have been linked to a range of neurological and psychiatric conditions, including AD, schizophrenia, and traumatic brain injury, which often reveal disrupted connectivity patterns that correlate with cognitive deficits or disease severity[13–16]. These network-level analyses provide a systems-level perspective on the manifestation of white matter disruptions across the brain, which enables researchers to move beyond isolated regional changes and towards a more integrated understanding of brain pathology[17].

In addition to traditional network analyses, autoencoder based methods have emerged as powerful tools to extract meaningful patterns from structural connectomes. Unsupervised approaches aim to uncover latent structure in the data without external labels, which makes them well-suited for dimensionality reduction, clustering, and learning low-dimensional manifolds that capture key sources of variation. For example, Zhang et al[18] showed that correlation-based feature selection on connectomes followed by k-means clustering differentiates healthy controls from schizophrenia patients. Such dimensionality reduction enables the modeling of intrinsic variance in high-dimensional data without labeled outcomes and serves as an effective feature extraction technique. Xing et al[19]. demonstrated that Riemann kernel PCA distinguishes healthy controls from individuals with obsessive-compulsive disorder, while Liu et al[20]. and Zhang et al[21]. proposed methods such as principal parcellation analysis and tensor network PCA to extract features predictive of traits like cognition, motor function, and sensory integration.

From a more machine learning–oriented perspective, autoencoders offer a flexible framework to learn low-dimensional latent representations directly from the data[22]. Parisot et al[23]. demonstrated that autoencoders outperform traditional techniques such as PCA and recursive feature elimination, when applied to vectorized connectomes, differentiates autism spectrum and AD from healthy controls. Newlin et al[24]. use conditional variational autoencoders (c-VAEs), which incorporate probabilistic representations to disentangle biological-relevant patterns from sources of technical variability in connectome data. However, a common limitation across these methods is that they model the latent space as purely continuous (Figure 1), which makes it difficult to capture inherently discrete sources of variation, such as acquisition site, protocol, or diagnostic subgroup. These discrete factors are often crucial for understanding the structure of neuroimaging datasets and may not be well represented by continuous embeddings alone.

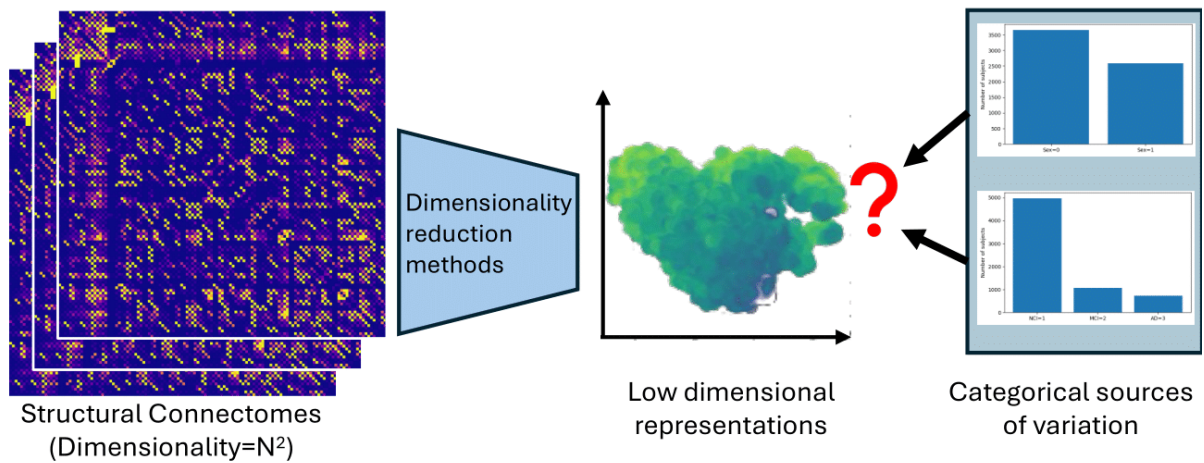

Fig 1. Traditional dimensionality reduction methods are useful for summarizing the high-dimensional structure of connectomes into a few continuous components that explain most of the variance. However, these methods are limited in their ability to capture intrinsic sources of discrete variation, such as sex or site difference, since they rely solely on continuous latent representations. This poses a challenge for interpreting and disentangling mixed sources of variability in structural connectome data.

Recent advances in machine learning now enable models to jointly learn both continuous and discrete latent spaces, allowing for richer representations that reflect the mixed nature of variation in complex datasets. Dupont et al[25]. showed that combining discrete latent variables alongside continuous ones can improve disentanglement and capture the inherent data variability more effectively. However, their approach relies heavily on a heuristic margin-hinge term in the losses, which encourages but does not guarantee balanced capacity between the two latent spaces, limiting precise control when such balance is critical in medical imaging. To address this, we propose joint VAE with staged capacity control, a novel framework that explicitly regulates the effective capacity of the continuous and discrete spaces during training. Our approach promotes balanced utilization of the two spaces with guaranteed capacity control. We evaluate these joint latent spaces in the context of structural connectomes, aiming to understand and characterize what the discrete component of the latent space captures. We compare our approach to traditional dimensionality reduction techniques such as PCA and standard VAEs followed by clustering, evaluating whether the discrete latent space learned through this joint modeling framework reliably captures meaningful structure in the data. Through this comparison, we assess whether modeling discrete variation explicitly offers a more robust and interpretable representation of the underlying connectome variability.

## 2. METHODS

### 2.1 Data

To investigate the applicability of hybrid continuous and discrete latent space models for structural connectomes, we curated a dataset from six major diffusion MRI studies focused on AD. To avoid bias from repeated measurements, we included only one scan per subject, despite the longitudinal nature of many of these studies. The final dataset consists of 955 connectomes from the Baltimore Longitudinal Study of Aging (BLSA)[26], 2486 from the Health and Aging Brain Study - Health Disparities (HABSHD)[27] study, 339 from the Wisconsin Registry for Alzheimer's Prevention (WRAP)[28], 1006 connectomes from the National Alzheimer's Coordinating Center (NACC)[29] study, 610 connectomes from a combined cohort of the Religious Orders Study(ROS), the Memory and Aging Project(MAP)[30] and the Minority Aging Research Study (MARS)[31].

Within each study, we defined a "site" as a unique combination of scanner location and acquisition protocol, resulting in 10 unique sites across the dataset. The cohort includes 3579 females and 2182 males, with clinical diagnoses spanning healthy control (n = 4338), mild cognitive impairment (MCI; n = 809), and Alzheimer's disease (AD; n = 614) and an age range of 22 to 101 years.

### 2.2 Preprocessing

We preprocessed the raw diffusion MRI scans with the PreQual[32] pipeline, which performs denoising and corrects for motion and eddy current artifacts to improve data quality. After PreQual, we estimated the fiber orientation distribution (FOD) at each voxel and performed whole brain tractography using MrTrix[33]. We seeded streamlines at the white matter–gray matter boundary, and we generated a total of 10 million streamlines with a probabilistic tracking algorithm [CITE]. To construct the connectomes, we parcellated the brain into regions using the SLANT[34] atlas and generated the connectome by counting the number of streamlines between each of the parcels. At every step of the preprocessing, we performed visual quality checks to make sure the processing was generating reasonable results following the procedure outlined by Kim et al.[35,36]

### 2.3 Joint-VAE

Traditional variational autoencoders[37] have an encoder: a neural network parameterized by $\emptyset$ which learns a function $q_\emptyset(z|x)$ to map the inputs x into a latent space z. Then the decoder, parameterized by $\theta$, learns a function $p_\theta(x|z)$ which reconstructs the inputs back from the latent space. While creating this mapping, the latent space is also encouraged to follow a prior distribution $p(z)$ which is typically a gaussian distribution. So, the loss of a traditional VAE then can be written as:

$$\mathcal{L}(\theta, \emptyset) = \mathbb{E}_{q_\emptyset(z|x)}[\log p_\theta(x|z)] - D_{KL}(q_\emptyset(z|x) \parallel p(z)) \qquad (1)$$

Where $\mathbb{E}_{q_\emptyset(z|x)}[\log p_\theta(x|z)]$ encourages the model to learn a good way to compress the data so that it can be built back from the latent space and $D_{KL}(q_\emptyset(z|x) \parallel p(z))$ encourages the learnt latent space to follow the prior distribution, which in this case is an $\mathcal{N}(0,1)$ gaussian distribution. We can increase the importance of following this prior distribution by weighting it by a factor $\beta$ which when greater than 1, which results in each in the dimensions of the learnt latent space to learn independent factors of variation in the input data giving rise to disentangled representations[38].

Dupont et al[25]. extended the $\beta$-VAE framework and showed that we can build joint discrete and continuous latent spaces in a principled way. According to their formulation, if $z_c$ is the continuous latent vector and $z_d$ is the discrete latent vector, the encoding can be defined as the joint posterior $q_\phi(z_c, z_d|x)$ under the decoder, the prior can be defined as $p(z_c, z_d)$ and the likelihood becomes $p_\theta(x|z_c, z_d)$ making the latent space jointly discrete and continuous. The loss from Equation 1 then becomes:

$$\mathcal{L}(\theta, \phi) = \mathbb{E}_{q_\phi(z_c, z_d|x)}[\log p_\theta(x|z_c, z_d)] - D_{KL}(q_\phi(z_c, z_d|x) \| p(z_c, z_d)) \quad (2)$$

Then, under the assumption that the priors and the posteriors are conditionally independent $q_\phi(z_c, z_d|x) = q_\phi(z_c|x) q_\phi(z_d|x)$ and $p(z_c, z_d) = p(z_c)p(z_d)$, and we can rewrite the KL divergence term, decomposing the optimization objective as follows:

$$D_{KL}(q_\phi(z_c, z_d|x) \| p(z_c, z_d)) = D_{KL}(q_\phi(z_c|x) \| p(z_c)) + D_{KL}(q_\phi(z_d|x) \| p(z_d)) \quad (3)$$

$$\mathcal{L}(\theta, \phi) = \mathbb{E}_{q_\phi(z_c, z_d|x)}[\log p_\theta(x|z_c, z_d)] - \beta|D_{KL}(q_\phi(z_c|x) \| p(z_c)) - C_z| - \beta|D_{KL}(q_\phi(z_d|x) \| p(z_d) - C_d)| \quad (4)$$

However, optimizing this objective directly often causes the model to rely primarily on the continuous latent space $z_c$, while underutilizing the discrete space $z_d$. Building on the idea that guiding the capacity of the KL divergence term can improve reconstruction quality and disentanglement[38], Dupont et al.[25] proposed gradually increasing the capacity of the for both latent spaces, allowing them to contribute at different rates during training. This controlled capacity scheduling strategy encourages more balanced use of the latent space. In our experiments we found that training the model with a capacitive loss as defined in Equation 4 resulted in the model often not using the discrete classes, suggesting sensitivity to the chosen margin-hinge hyperparameters $C_z$ and $C_d$.

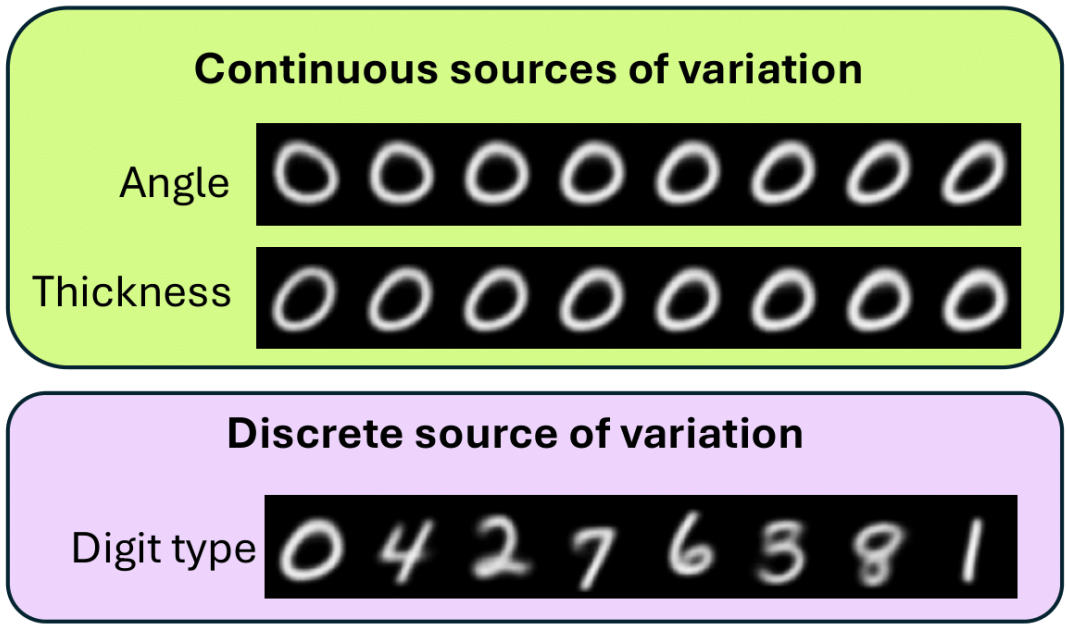

Fig 2. Recent advances in machine learning enable the construction of joint continuous and discrete latent spaces using a Joint Variational Autoencoder (Joint-VAE). For example, in the MNIST dataset, the continuous latent variables can capture digit-specific variations such as thickness or rotation, while the discrete variables encode digit identity, all learned in an unsupervised manner. Inspired by this, we apply the same framework to structural connectomes, aiming to disentangle the underlying sources of variability and reduce the high-dimensional connectome data into a meaningful latent space in a data-driven way.

To address this, we propose a joint-VAE with staged capacity control. We replace the learned continuous latent parameters with a convex combination of their learned values and the standard Gaussian prior parameters. Specifically, we set $\mu' = \lambda\mu$ and $\log \sigma'^2 = \lambda \log \sigma^2 + (1 - \lambda) \log 1 = \lambda \log \sigma^2$, where $\lambda$ increases linearly from 0 to 1 over a fixed number of iterations. This modification guarantees that the effective capacity transitions smoothly from zero (i.e., matching the prior at $\lambda = 0$) to full capacity ($\lambda = 1$). Subsituting these scaled parameters into the KL divergences terms yields Equation 5, providing a principled mechanism for controlling the balance between continuous and discrete capacity.

$$\widetilde{D_{KL}}\big(q_\emptyset\,(z_c|x)\,\|\,p(z_c)\big) = \tfrac{1}{2}(\sigma^{2\lambda} + \lambda^2\mu^2 - 1 - \lambda\log\sigma^2). \quad (5)$$

$$\mathcal{L}(\theta,\emptyset) = \mathbb{E}_{q_\emptyset(z_c,z_d|x)}[\log p_\theta(x|z_c,z_d)] - \beta\widetilde{D_{KL}}\big(q_\emptyset(z_c|x)\,\|\,p(z_c)\big) - \beta D_{KL}\big(q_\emptyset(z_d|x)\,\|\,p(z_d)\big) \quad (6)$$

At early epochs, the KL penalty on the continuous latent space is minimal, encouraging the model to rely more heavily on the discrete space. As training progresses and the annealing factor λ gradually increases to 1, the full KL regularization is applied, promoting balanced use of both latent spaces. To better understand the behavior of the hybrid latent space, we reproduced the experiment conducted by Dupont et al[25]. with our modified objective in (equation 6) on the MNIST dataset. This simple dataset allows clear visualization of how the model separates discrete and continuous factors. We used a 3-dimensional continuous latent space and a 10-class discrete latent space and observed that the continuous dimensions captured smooth variations such as stroke angle and thickness, while the discrete latent captured digit identity (Figure 2). For example, varying one continuous dimension smoothly altered the stroke angle, while another controlled thickness. In contrast, changing the discrete variable resulted in different digits being generated, while keeping stroke style fixed, which demonstrates that the model disentangles the factors as expected.

### 2.4 Model architecture

In our work, we design a VAE with a joint continuous and discrete latent space tailored for structural connectomes (Figure 3). The encoder takes the flattened upper triangular portion of the connectome as input and passes it through four fully connected layers, each followed by a ReLU activation. From the final encoder layer, we obtain three outputs: the mean vector ($\mu$), the log-variance vector ($\log\sigma^2$), and the logits vector $\alpha$, which parameterizes the discrete latent distribution. The continuous latent space $z_c$ is sampled using the standard VAE reparameterization trick, where each dimension follows a Gaussian distribution $q_\emptyset(z_{c_i}|x) = \mathcal{N}(\mu_i, \sigma^2_i)$. The discrete latent space $z_d$ models 10 categorical classes. The $\alpha$ vector represents unnormalized log-probabilities for each class, and we approximate sampling from this categorical distribution using the Gumbel-Softmax trick[39], which allows differentiable sampling during training. We assume a uniform prior $p(z_d)$ over the discrete classes, implemented as a uniform Gumbel-Softmax distribution. We trained the model for 3000 epochs with a batch size of 512, using $\beta = 5$ and linearly increasing λ from 0 to 1 over the first 3500 iterations.

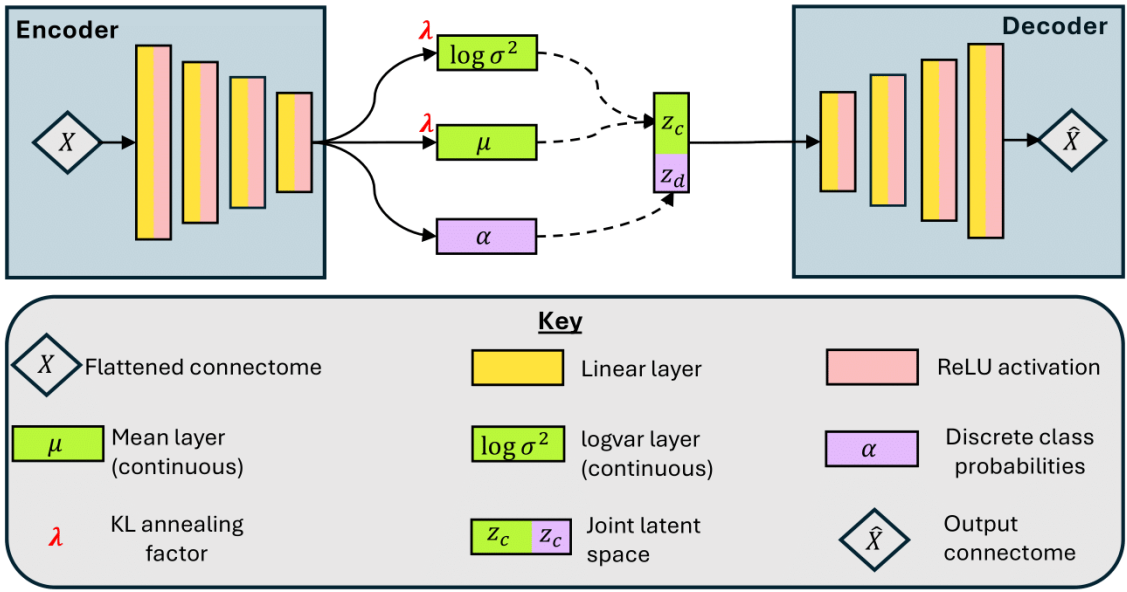

Fig 3. We implement a joint-VAE that learns to separate continuous and discrete variances from the input connectomes ($X$). The encoder outputs three components: (1) a mean and log-variance vector, which are used to sample the continuous latent space ($z_c$) via the standard VAE reparameterization trick; and (2) a vector of class probabilities $\alpha$ which is reparametrized using the Gumbel-SoftMax trick and argmaxed to yield the discrete latent variable ($z_d$). The continuous KL divergence is scaled by an annealing factor λ that linearly increases from 0 to 1 over a set number of iterations. The final latent representation, formed by concatenating $z_c$ and $z_d$, is passed to the decoder to reconstruct the output connectome ($\hat{X}$).

## 2.5 Experimental design

To assess whether the discrete latent space captures meaningful variation related to known metadata, we trained nine separate models with the number of discrete classes ranging from 2 to 10. For each trained model, we encoded all connectomes and analyzed the learned discrete class assignments. As baselines, we compared against (i) PCA followed by k-means clustering on the top nine components (chosen using the elbow method), and (ii) a standard VAE with continuous latent space followed by k-means. We quantified the alignment of clustering results with site labels using the Adjusted Rand Index (ARI)[40], which measures the similarity between the clusters learned by the model and the ground truth site labels and computed confidence intervals through 1000 bootstrap resampling. ARI is particularly useful in this context, as it accounts for agreement across different numbers of clusters.

## 3. RESULTS

We analyzed the discrete latent space by visualizing the learned embeddings with t-SNE, coloring each point by its assigned discrete class and comparing it to the metadata. The resulting clusters aligned strongly with site labels, which suggests that the discrete space effectively captures site-related variation (Figure 4). While the baseline could partially recover site structure, we found that the discrete latent space learned by our joint VAE preserved site-specific clusters more clearly. For instance, the largest site in our dataset, HABSHD, remained largely intact as a single cluster in the discrete space, whereas PCA+k-means split it across multiple clusters.

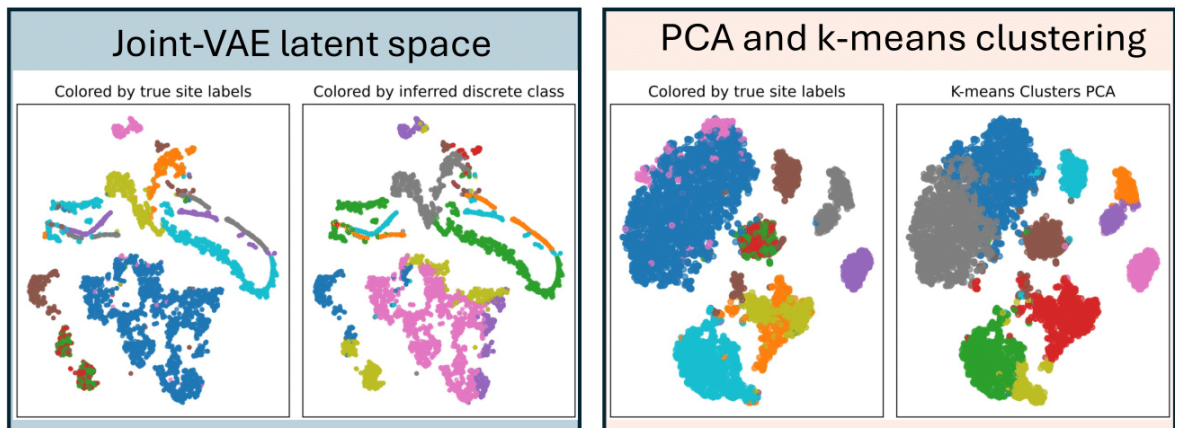

Fig 4. (Left) The continuous latent representations from our joint-VAE model are colored by the discrete classes learned by the model. These discrete classes align closely with the true site labels, indicating that the model effectively captures site-related variation. (Right) PCA is applied to the original connectomes, followed by k-means clustering on the principal components. Compared to our model, PCA + k-means fails to recover the same level of site separation, suggesting that the joint-VAE captures intrinsic site-related structure that is not readily uncovered by standard dimensionality reduction and clustering methods.

Quantitatively, our results show that, while all methods perform similarly with a small number of clusters, the performance of the PCA and standard VAE baselines deteriorates as k increases. In contrast, the joint VAE remains stable and continues to capture meaningful site-related variation (Figure 5). At k = 10, which corresponds to the true number of sites, the ARI achieved by the joint VAE is significantly higher than both baselines ($p \ll 0.05$), with significance assessed using a t-test on the bootstrapped samples. This indicates that the discrete latent space learned by the joint VAE captures site-specific structure more effectively than traditional dimensionality reduction methods.

## 4. DISCUSSION AND CONCLUSION

Our results demonstrate that a joint continuous–discrete latent representation provides a stable and interpretable summary of structural connectomes. Notably, the discrete component consistently captured site-related variation and preserved coherent clusters as the number of assumed classes increased, outperforming PCA+k-means and VAE+k-means baselines, particularly at higher k. This effect, quantified using ARI with bootstrapped confidence intervals, highlights the advantage of explicitly modeling discrete variation. At k=10, which matches the true number of acquisition sites, the discrete latent assignments aligned most strongly with site labels, which indicates that the model successfully isolated a key source of technical variability inherent to multi-site diffusion MRI datasets.

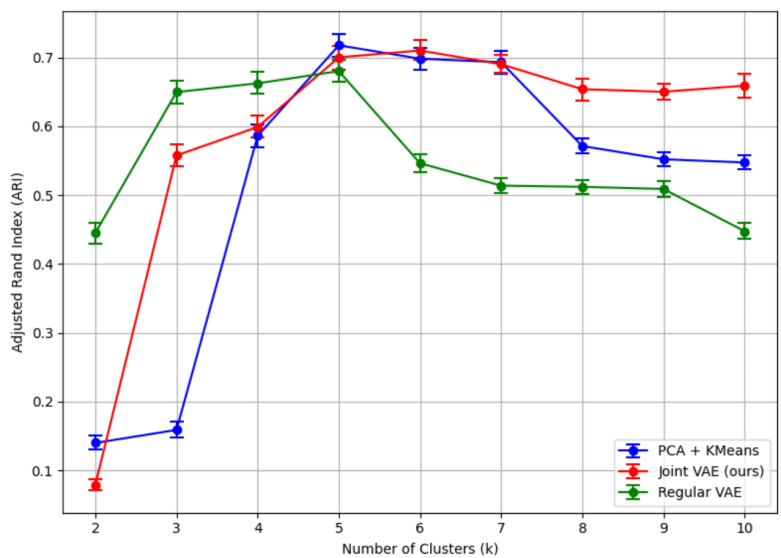

Fig 5. We assess how well the discrete latent variables from our joint-VAE model capture site-related variation by computing the Adjusted Rand Index (ARI) between site labels and unsupervised cluster assignments. At lower cluster numbers, our method performs comparably to the PCA + k-means baseline but maintains significantly greater stability as the number of clusters increases. Notably, baseline performance declines beyond k = 7, whereas our method continues to perform well. At k = 10, the true number of sites, our method significantly outperforms both baselines ($p < 0.05$), indicating that the discrete latent space effectively captures site-specific structure. Confidence intervals and p-values were obtained via 1000 bootstrap resamples.

From a modeling perspective, our use of KL annealing and capacity scheduling addressed the common failure mode in which the continuous latent space dominates while the discrete space is ignored. By gradually increasing the KL penalty on the continuous dimensions, the model is encouraged early in training to make use of the discrete codes, yielding a more balanced distribution of variance across the latent space. In practice, these discrete codes can serve as harmonization covariates or quality-control signals. Xu et al[41]. showed that traditional harmonization like mean-shift, Combat and CycleGAN methods showed poor performance on structural connectomes and their derived metrics. Newlin et al[24]. demonstrated that connectomes could be harmonized using a conditional VAE framework, but this approach requires access to site labels. In contrast, our joint VAE automatically recovers site-related clusters in a purely data-driven manner, assuming the site signal is sufficiently strong, which is often the case in medical imaging data from multi-site datasets[42,43]. As imaging studies increasingly scale to large, heterogeneous cohorts, this framework provides a robust method to disentangle and explain technical variation in the discrete space, which enables downstream modeling of biological variance in the continuous space, potentially with added supervision.

While our study demonstrates the utility of hybrid latent spaces in disentangling site-related variation, there are several important directions for further investigation. Our dataset comprises only Alzheimer's-related cohorts and ten acquisition protocols; applying this framework to additional diseases, scanner types, and acquisition settings would assess its generalizability. Moreover, although the discrete latent space clearly captures site-related structure, further analysis is needed to determine how effectively the continuous space retains meaningful biological variation such as age or cognitive status after adjusting for site effects. Exploring these relationships could offer deeper insights into the separation of technical and biological signals within the latent space. Future work could also examine how incorporating contrastive learning or domain-adversarial objectives through supervision to further enhance disentanglement and robustness across datasets.

## 5. ACKNOWLEDGEMENTS

This work was conducted in part using the resources of the Advanced Computing Center for Research and Education at Vanderbilt University, Nashville, TN, and was supported by NIH grant R01EB017230 (PI: Landman). The Vanderbilt Institute for Clinical and Translational Research (VICTR) is supported by the National Center for Advancing Translational Sciences (NCATS) Clinical Translational Science Award (CTSA) Program under award number UL1TR002243. This work was supported by the Alzheimer's Disease Sequencing Project Phenotype Harmonization Consortium (ADSP-PHC) that is funded by NIA (U24 AG074855, U01 AG068057 and R01 AG059716). This work was also supported by the National Cancer Institute (NCI) grants R01 CA253923 and R01 CA275015

This research was supported in part by the Intramural Research Program of the National Institutes of Health (NIH). The contributions of the NIH author(s) were made as pasrt of their official duties as NIH federal employees, are in compliance with agency policy requirements, and are considered Works of the United States Government. However, the findings and conclusions presented in this paper are those of the author(s) and do not necessarily reflect the views of the NIH or the U.S. Department of Health and Human Services.

The BLSA dataset was supported by the Intramural Research Program of the National Institute on Aging, NIH. Data from the Wisconsin Registry for Alzheimer's Prevention (WRAP) was supported by NIA grants AG021155, AG0271761, AG037639, and AG054047.

We gratefully acknowledge the efforts of the HABS-HD MPIs: Sid E. O'Bryant, Kristine Yaffe, Arthur Toga, Robert Rissman, and Leigh Johnson, as well as the HABS-HD Investigators: Meredith Braskie, Kevin King, James R. Hall, Melissa Petersen, Raymond Palmer, Robert Barber, Yonggang Shi, Fan Zhang, Rajesh Nandy, Roderick McColl, David Mason, Bradley Christian, Nicole Phillips, Stephanie Large, Joe Lee, Badri Vardarajan, Monica Rivera Mindt, Amrita Cheema, Lisa Barnes, Mark Mapstone, Annie Cohen, Amy Kind, Ozioma Okonkwo, Raul Vintimilla, Zhengyang Zhou, Michael Donohue, Rema Raman, Matthew Borzage, Michelle Mielke, Beau Ances, Ganesh Babulal, Jorge Llibre-Guerra, Carl Hill, and Rocky Vig. Research related to HABS-HD was supported by the National Institute on Aging of the National Institutes of Health under award numbers R01AG054073, R01AG058533, R01AG070862, P41EB015922, and U19AG078109.

Data contributed from the ROS/MAP/MARS studies were supported by grants from the National Institute on Aging: R01AG017917, P30AG10161, P30AG072975, R01AG022018, R01AG056405, UH2NS100599, UH3NS100599, R01AG064233, R01AG15819, and R01AG067482, along with support from the Illinois Department of Public Health (Alzheimer's Disease Research Fund). These data are available at www.radc.rush.edu.

Data were also provided in part by OASIS-4: Clinical Cohort, led by Principal Investigators T. Benzinger, L. Koenig, and P. LaMontagne.

We also acknowledge the National Alzheimer's Coordinating Center (NACC) database, which is funded by NIA/NIH Grant U24AG072122. Data were contributed by NIA-funded Alzheimer's Disease Research Centers (ADRCs), including but not limited to: P30AG062429 (PI: James Brewer), P30AG066468 (PI: Oscar Lopez), P30AG062421 (PI: Bradley Hyman), P30AG066509 (PI: Thomas Grabowski), P30AG066514 (PI: Mary Sano), P30AG066530 (PI: Helena Chui), P30AG066507 (PI: Marilyn Albert), P30AG066444 (PI: John Morris), and numerous others listed in full above.